\documentclass[12pt]{article}
\usepackage{amsmath,amssymb,amsthm,amsxtra,overpic,bbm,bm,epsfig,ulem,subfigure}
\usepackage{color}
\usepackage{hyperref}
\usepackage{mathrsfs}
\usepackage{graphicx}
\usepackage{comment}
\usepackage{float}
\usepackage{cite}
\usepackage{url}
\usepackage{slashed,stmaryrd}

\def\thefootnote{\fnsymbol{footnote}}

\begin{document}

\vspace{0.2cm}

\begin{center}
{\large\bf Neutrino cuboid for normal mass ordering and tribimaximal flavor mixing}
\end{center}

\vspace{0.2cm}

\begin{center}
{\bf Zhi-zhong Xing$^{1,2,3}$}
\footnote{E-mail: xingzz@ihep.ac.cn}
\\
{\small $^{1}$ Institute of High Energy Physics, Chinese Academy of
Sciences, Beijing 100049, China \\
$^{2}$ School of Physical Sciences,
University of Chinese Academy of Sciences, Beijing 100049, China \\
$^{3}$ Center of High Energy Physics, Peking University, Beijing 100871, China}
\end{center}

\vspace{2cm}
\begin{abstract}
Given the latest JUNO implication for normal neutrino mass ordering,
we parametrize three neutrino masses in a flavor cuboid:
$m^{}_1 = m^{}_0 \sin\xi$, $m^{}_2 = m^{}_0 \cos\xi \sin\zeta$ and
$m^{}_3 = m^{}_0 \cos\xi \cos\zeta$. We find that this cuboid is able
to accommodate both neutrino mass degeneracy and tribimaximal flavor
mixing in its cubic limit with $\xi^{}_* = \arctan\left(1/\sqrt{2}\right)
\simeq 35.26^\circ$ and $\zeta^{}_* = 45^\circ$. Assuming
$\theta^{}_{12} = \xi$ and $\theta^{}_{23} = \zeta$ for the
two large angles of neutrino oscillations and
expanding them around $\xi^{}_*$ and $\zeta^{}_*$, we propose a
viable ansatz which predicts a normal but nearly degenerate neutrino
mass spectrum and a nearly tribimaximal neutrino mixing pattern.
Testing the achieved correlation among $\xi^{}_* - \theta^{}_{12}$,
$\zeta^{}_* - \theta^{}_{23}$ and $\Delta m^2_{21}/\Delta m^2_{31}$
will provide a smoking gun for the validity of this ansatz.
\end{abstract}

\newpage

\def\thefootnote{\arabic{footnote}}
\setcounter{footnote}{0}
\setcounter{equation}{0}
\setcounter{figure}{0}

In the standard model (SM) of particle physics, all the
fundamental charged fermions of the same electric charge have a normal
mass hierarchy ($m^{}_u \ll m^{}_c \ll m^{}_t$ for quarks of
$Q = +2/3$, $m^{}_d \ll m^{}_s \ll m^{}_b$ for quarks of $Q = -1/3$
and $m^{}_e \ll m^{}_\mu \ll m^{}_\tau$ for charged leptons of
$Q = -1$), but whether the free neutrinos $\nu^{}_1$, $\nu^{}_2$
and $\nu^{}_3$ corresponding to $\nu^{}_e$, $\nu^{}_\mu$ and
$\nu^{}_\tau$ flavors also possess the $m^{}_1 < m^{}_2 < m^{}_3$
ordering or not has been an open question~\cite{Xing:2019vks}
\footnote{When addressing the neutrino mass ordering issue, one has taken
the standard form of leptonic weak charged-current interactions in the mass
bases of three charged leptons ($e$, $\mu$, $\tau$) and three active
neutrinos ($\nu^{}_1$, $\nu^{}_2$, $\nu^{}_3$), and fixed the pattern of the
$3\times 3$ lepton flavor mixing matrix $U$ with its three columns being
$U^{}_{\alpha 1}$, $U^{}_{\alpha 2}$ and $U^{}_{\alpha 3}$ (for
$\alpha = e, \mu, \tau$). As a result, a definite and unique ordering
among $m^{}_1$, $m^{}_2$ and $m^{}_3$ will be experimentally determined.
Reordering $\nu^{}_i$ and thus $m^{}_i$ (for $i = 1, 2, 3$) is certainly
allowed from a mathematical point of view, but in this case one has to
simultaneously reordering the columns
of $U$ to keep the relevant physics unchanged~\cite{Xing:2017cwb}.}.
To answer this question, the {\it conventional} idea is to
measure the signs of the neutrino mass-squared differences
$\Delta m^2_{ij} \equiv m^2_i - m^2_j$ (for $i, j = 1, 2, 3$) in solar,
atmospheric and long-baseline accelerator neutrino oscillation experiments
with the help of heliacal or terrestrial matter
effects~\cite{ParticleDataGroup:2026}. So far $\Delta m^2_{21} >0$
has been determined in this way, and $\Delta m^2_{21} \ll
\left|\Delta m^2_{31}\right| \simeq \left|\Delta m^2_{32}\right|$
has been established too. These measurements imply that $\Delta m^2_{31}$
and $\Delta m^2_{32}$ must have the same sign and thus their positivity
(or negativity) points to the normal $m^{}_1 < m^{}_2 < m^{}_3$ case (or
the inverted $m^{}_3 < m^{}_1 < m^{}_2$ case). In comparison, an
{\it unconventional} method to determine the sign of $\Delta m^2_{31}$ is
to accurately observe the medium-baseline oscillation behaviors of reactor
antineutrinos in the JUNO experiment~\cite{Zhan:2008id,Zhan:2009rs,Li:2013zyd},
which is not only free from CP violation but also insensitive to terrestrial
matter effects. The first JUNO hint of $\Delta m^2_{31} >0$ has recently been
reported at the $2.3~\sigma$ confidence level~\cite{Wang:2026}. Given the
very fact that the sign of $\Delta m^2_{31}$ is intrinsically a
{\it black-and-white} quantity instead of a usual observable whose value may
fluctuate due to various experimental uncertainties, we expect that the JUNO
measurement should serve as a quite strong indication of normal neutrino
mass ordering.
\begin{figure}[t]
\begin{center}
\vspace{0.17cm}
\includegraphics[width=5.35cm]{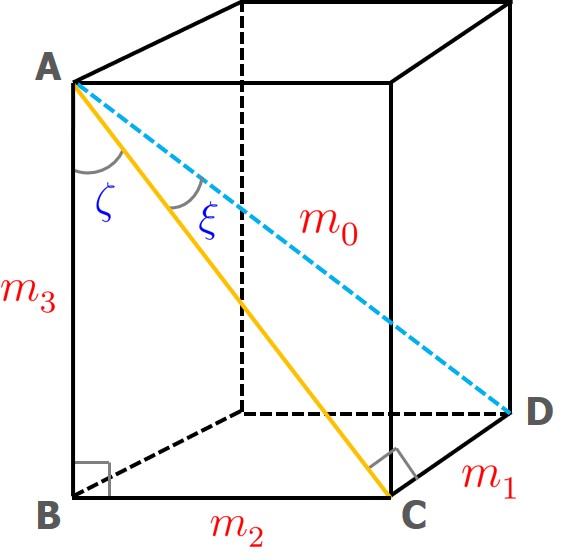}
\vspace{-0.22cm}
\caption{A geometrical parametrization of the three neutrino masses based
on a flavor cuboid for the $m^{}_1 < m^{}_2 < m^{}_3$ ordering,
where $m^{}_0 = \displaystyle \sqrt{\displaystyle m^2_1 + m^2_2 + m^2_3}
\hspace{0.05cm}$, $\zeta < 45^\circ$ and $\cot \xi \sin\zeta > 1$.}
\label{mass-plot}
\end{center}
\end{figure}

In the lack of a convincing determination of the absolute neutrino mass scale
$m^{}_0$~\cite{ParticleDataGroup:2026}, this short note is intended to take
the JUNO result seriously and advocate a geometrical parametrization of the normal
neutrino mass spectrum as illustrated by the flavor cuboid in
Fig.~\ref{mass-plot}:
\begin{eqnarray}
m^{}_1 \hspace{-0.2cm} & = & \hspace{-0.2cm} m^{}_0 \sin\xi \; ,
\nonumber \\
m^{}_2 \hspace{-0.2cm} & = & \hspace{-0.2cm} m^{}_0 \cos\xi
\sin\zeta \; ,
\nonumber \\
m^{}_3 \hspace{-0.2cm} & = & \hspace{-0.2cm} m^{}_0 \cos\xi
\cos\zeta \; ,
\label{1}
\end{eqnarray}
where $m^{}_0 = \displaystyle \sqrt{\displaystyle
m^2_1 + m^2_2 + m^2_3}$ characterizes the length of the space diagonal,
$\xi$ and $\zeta$ are two intersection angles lying in the lower octant
of the first quadrant and satisfying the precondition
$\cot\xi \sin\zeta > 1$ to guarantee $m^{}_3 > m^{}_2 > m^{}_1$.
So $\xi \to 0$ will result in $m^{}_1 \to 0$, an interesting
extreme which makes the cuboid reduced to a two-dimensional rectangle and
thus allows one to determine $m^{}_2$ and $m^{}_3$ in terms of
the observed values of $\Delta m^2_{21}$ and
$\Delta m^2_{31}$~\cite{Xing:2019vks}. But we find that it is theoretically
more nontrivial and phenomenologically more intriguing
to consider the other extreme
\footnote{Given definite values of the two independent neutrino mass-squared
differences and the three flavor mixing angles extracted from a variety of
neutrino oscillation experiments, it is well known that the sum of three
neutrino masses and the effective electron neutrino mass terms of both the
beta decays and the neutrinoless double-beta decays tend to take their
maxima when the three neutrinos have a nearly degenerate mass spectrum
and the unknown CP-violating phases are allowed to arbitrarily
vary~\cite{ParticleDataGroup:2026}. A near neutrino mass degeneracy is
therefore favored from the point of view of maximal experimental
observability, and it also implies a profound difference between the origins
of three neutrino masses and three charged-lepton masses from a theoretical
angle of view.} ---
the neutrino mass degeneracy
$m^{}_1 = m^{}_2 = m^{}_3 = m^{}_0/\sqrt{3}$ under the conditions
\begin{eqnarray}
\xi^{}_* = \arctan\left(\frac{1}{\sqrt{2}} \right) \simeq 35.26^\circ \; ,
\quad \zeta^{}_* = 45^\circ \; ,
\label{2}
\end{eqnarray}
which are completely compatible with the values
of $\theta^{}_{12}$ and $\theta^{}_{23}$ predicted by the well-known
tribimaximal lepton mixing pattern in a class of flavor symmetry
models~\cite{Harrison:2002er,Xing:2002sw,He:2003rm}. In this special case,
the flavor cuboid in Fig.~\ref{mass-plot} is reduced to a cube~\cite{Lee:2006pr}.
Assuming $\theta^{}_{12} = \xi$ and $\theta^{}_{23} = \zeta$ for the
two large angles of neutrino oscillations and expanding them
around $\xi^{}_*$ and $\zeta^{}_*$ respectively, we are going to propose a
viable ansatz which predicts a normal but nearly degenerate
neutrino mass spectrum and a nearly tribimaximal flavor mixing pattern.
A careful test of the achieved correlation among $\xi^{}_* - \theta^{}_{12}$,
$\zeta^{}_* - \theta^{}_{23}$ and $\Delta m^2_{21}/\Delta m^2_{31}$
will provide a smoking gun for the validity of this ansatz,
especially in the precision JUNO, DUNE and Hyper-Kamiokande neutrino
oscillation experiments.

Let us start with the standard parametrization of the $3\times 3$ unitary
lepton flavor mixing matrix $U$ that has been advocated by the Particle Data
Group~\cite{ParticleDataGroup:2026},
\begin{eqnarray}
U = \left( \begin{matrix} c^{}_{12} c^{}_{13} & s^{}_{12}
c^{}_{13} & s^{}_{13} e^{-{\rm i}\delta} \cr
-s^{}_{12} c^{}_{23} - c^{}_{12} s^{}_{13} s^{}_{23} e^{{\rm i}\delta}
& c^{}_{12} c^{}_{23} -
s^{}_{12} s^{}_{13} s^{}_{23} e^{{\rm i}\delta} & c^{}_{13} s^{}_{23} \cr
s^{}_{12} s^{}_{23} - c^{}_{12} s^{}_{13} c^{}_{23} e^{{\rm i}\delta}
& -c^{}_{12} s^{}_{23} - s^{}_{12} s^{}_{13} c^{}_{23} e^{{\rm i}\delta}
& c^{}_{13} c^{}_{23}
\cr \end{matrix} \right)
\left( \begin{matrix}
e^{{\rm i}\rho} & 0 & 0 \cr
0 & e^{{\rm i}\sigma} & 0 \cr
0 & 0 & 1 \cr\end{matrix} \right) \; ,
\label{3}
\end{eqnarray}
in which $c^{}_{ij} \equiv \cos\theta^{}_{ij}$ and $s^{}_{ij} \equiv
\sin\theta^{}_{ij}$ with $\theta^{}_{ij}$ lying in the first quadrant
(for $ij = 12, 13, 23$), $\delta$ denotes the CP-violating phase
responsible for CP violation in neutrino oscillations, while $\rho$ and
$\sigma$ are the other two CP-violating phases associated with the
lepton-number-violating processes such as the neutrinoless double-beta
decays. The latest JUNO measurement offers
\begin{eqnarray}
&& \sin^2\theta^{}_{12} = 0.3036 \pm 0.0064 \; ,
\nonumber \\
&& \Delta m^2_{21} = \left(7.388 \pm 0.078\right) \times 10^{-5} ~
{\rm eV}^2 \; , \hspace{0.1cm}
\nonumber \\
&& \Delta m^2_{31} = \left(2.509^{+0.027}_{-0.025}\right) \times
10^{-3} ~ {\rm eV}^2 \; ,
\label{4}
\end{eqnarray}
at the $1~\sigma$ confidence level, where the value of $\Delta m^2_{31}$
is obtained from a combination of the JUNO and Daya Bay precision
data~\cite{Wang:2026,JUNO:2025gmd}. In addition, a
recent global analysis of the available atmospheric and accelerator
neutrino oscillation data gave the best-fit result of $\theta^{}_{23}$
and its $1~\sigma$ interval~\cite{Capozzi:2025wyn,Esteban:2024eli}:
\begin{eqnarray}
&& \sin^2\theta^{}_{23} = 0.473^{+0.023}_{-0.013} \quad
({\rm Capozzi ~{\it et~al}}) \; ,
\nonumber \\
&& \sin^2\theta^{}_{23} = 0.470^{+0.017}_{-0.014} \quad
({\rm Esteban ~{\it et~al}}) \; ,
\label{5}
\end{eqnarray}
in the $\Delta m^2_{31} >0$ case. We are therefore left with the
following two important observations.
\begin{itemize}
\item     The results
$\theta^{}_{12} = 33.44^{+0.39^\circ}_{-0.40^\circ}$
and $\theta^{}_{23} = 43.45^{+1.32^\circ}_{-0.74^\circ}$
(or $43.28^{+0.98^\circ}_{-0.80^\circ}$) extracted from Eqs.~(\ref{4}) and
(\ref{5}) indicate that these two large flavor mixing angles deviate only
slightly from their corresponding tribimaximal mixing values
$\xi^{}_*$ and $\zeta^{}_*$. To be more explicit, $\varepsilon^{}_\xi \equiv
\xi^{}_* - \theta^{}_{12} \simeq 1.8^\circ$ is quite accurate,
while $\varepsilon^{}_\zeta \equiv \zeta^{}_* - \theta^{}_{23} \sim 1.6^\circ$
involves some much larger uncertainties. As a result, one may safely
expand the sine and cosine functions of $\theta^{}_{12}$ and $\theta^{}_{23}$
around their respective $\xi^{}_*$ and $\zeta^{}_*$ in powers of the
small parameters $\varepsilon^{}_\xi$ and $\varepsilon^{}_\zeta$.

\item     The intersection angles $\xi$ and $\zeta$ in Eq.~(\ref{1}),
which characterize the degree of hierarchy or degeneracy of three neutrino
masses, are constrained by the observed ratio of $\Delta m^2_{21}$ to
$\Delta m^2_{31}$:
\begin{eqnarray}
\eta \equiv \frac{\Delta m^2_{21}}{\Delta m^2_{32}} =
\frac{\sin^2\zeta - \tan^2\xi}{\cos^2\zeta - \tan^2\xi} \simeq
2.96 \times 10^{-2} \; .
\label{6}
\end{eqnarray}
Such a nonzero but very small $\eta$ implies that the extreme
neutrino mass hierarchy with $\xi = 0$ or equivalently
$\zeta = \arctan\left(\sqrt{\eta}\right) \simeq 9.76^\circ$ is
phenomenologically allowed, but the extreme neutrino mass degeneracy with
$\xi = \xi^{}_*$ and $\zeta = \zeta^{}_*$ is apparently forbidden.
\end{itemize}
Of course, a full determination of or a stringent constraint on the neutrino mass
spectrum requires some further inputs from those non-oscillation experiments
regarding the massive neutrinos, such as precision measurements of the sum of
three neutrino masses in cosmology or the beta decays and neutrinoless
double-beta decays in nuclear physics~\cite{ParticleDataGroup:2026}.

Here we propose a viable phenomenological ansatz of the neutrino mass spectrum
by just assuming $\xi = \theta^{}_{12}$ and $\zeta = \theta^{}_{23}$ for
Eq.~(\ref{1}), and expanding them around $\xi^{}_*$ and $\zeta^{}_*$ in terms of
$\varepsilon^{}_\xi$ and $\varepsilon^{}_\zeta$ respectively. Our key conjecture is
that the observed nearly tribimaximal lepton mixing pattern might be intrinsically
related to a near degeneracy of three neutrino masses
\footnote{A similar conjecture was first made by Fritzsch and the present author
in Ref.~\cite{Fritzsch:1995dj}, to link a nearly democratic lepton flavor mixing
pattern with a nearly degenerate neutrino mass spectrum.},
leading us to a nearly cubic flavor cuboid which will soon be tested in both
the JUNO reactor antineutrino oscillation experiment and those long-baseline
accelerator neutrino oscillation experiments.

After some straightforward calculations along the above line of thought, we simply
arrive at the following analytical approximations for the three neutrino masses:
\begin{eqnarray}
&& m^{}_1 \simeq \frac{1}{\sqrt 3} \left(1 - \sqrt{2} \hspace{0.08cm}
\varepsilon^{}_\xi\right) m^{}_0 \; ,
\nonumber \\
&& m^{}_2 \simeq \frac{1}{\sqrt 3} \left(1 + \frac{1}{\sqrt 2} \hspace{0.08cm}
\varepsilon^{}_\xi - \varepsilon^{}_\zeta\right) m^{}_0 \; ,
\nonumber \\
&& m^{}_3 \simeq \frac{1}{\sqrt 3} \left(1 + \frac{1}{\sqrt 2} \hspace{0.1cm}
\varepsilon^{}_\xi + \varepsilon^{}_\zeta\right) m^{}_0 \; , \hspace{0.1cm}
\label{7}
\end{eqnarray}
from which the three neutrino mass-squared differences are found to be
\begin{eqnarray}
&& \Delta m^2_{21} \simeq \frac{1}{3} \left(3\sqrt{2} \hspace{0.05cm}
\varepsilon^{}_\xi - 2 \hspace{0.08cm}\varepsilon^{}_\zeta \right) m^2_0 \; ,
\nonumber \\
&& \Delta m^2_{31} \simeq \frac{1}{3} \left(3\sqrt{2} \hspace{0.05cm}
\varepsilon^{}_\xi + 2 \hspace{0.08cm}\varepsilon^{}_\zeta\right) m^2_0 \; ,
\hspace{0.3cm}
\nonumber \\
&& \Delta m^2_{32} \simeq \frac{4}{3} \hspace{0.05cm}
\varepsilon^{}_\zeta \hspace{0.05cm} m^2_0 \; .
\label{8}
\end{eqnarray}
Some immediate comments on Eqs.~(\ref{7}) and (\ref{8}) are in order.
\begin{itemize}
\item     Given that both $\varepsilon^{}_\xi$ and $\varepsilon^{}_\zeta$
are positive, we see that the realistic flavor cuboid shown in Fig.~\ref{mass-plot}
originates from an extreme flavor cube ($m^{}_1 = m^{}_2 = m^{}_3 = m^{}_0/\sqrt{3}$)
by properly lengthening its height $m^{}_3$, shortening its width $m^{}_1$ and
adjusting its length $m^{}_2$ in our ansatz to get $m^{}_1 < m^{}_2 < m^{}_3$.
The flavor cube as a striking flavor symmetry limit (e.g., an $S^{}_3$ permutation
symmetry~\cite{Jora:2006dh,Xing:2010iu,Zhou:2011nu,Xing:2015fdg}) is therefore
the reasonable starting point for possible model building, and it is the
appropriate flavor symmetry breaking effects that simultaneously cause the
observed neutrino mass spectrum and flavor mixing pattern, including leptonic
CP violation.

\item     The small splitting between $m^{}_2$ and $m^{}_3$ is characterized by
the $\mu$-$\tau$ symmetry breaking parameter $\varepsilon^{}_\zeta$, which describes
to what extent the actual value of $\theta^{}_{23}$ deviates from $\zeta^{}_*
= 45^\circ$. In particular, the situation that $\theta^{}_{23}$ is located in
the lower octant of the first quadrant is closely correlated with normal
neutrino mass ordering.
Such interesting observations can not only help a lot for building a specific
flavor symmetry model, but also make it possible to experimentally
test this phenomenological ansatz in the near future.

\item     In the approximation made above for our ansatz, the expression of $\eta$
in Eq.~(\ref{6}) can be reduced to
\begin{eqnarray}
\eta = \frac{\sin^2\theta^{}_{23} - \tan^2\theta^{}_{12}}
{\cos^2\theta^{}_{23} - \tan^2\theta^{}_{12}}
\simeq \frac{3 \hspace{0.05cm} \varepsilon^{}_\xi - \sqrt{2} \hspace{0.08cm}
\varepsilon^{}_\zeta}{3 \hspace{0.05cm} \varepsilon^{}_\xi + \sqrt{2}
\hspace{0.08cm} \varepsilon^{}_\zeta} \; ,
\label{9}
\end{eqnarray}
indicating that there should exit an appreciable cancellation in the numerator of
$\eta$ to assure its smallness. It is quite easy to achieve the ratio
$\varepsilon^{}_\xi/\varepsilon^{}_\zeta \simeq \sqrt{2}\left(1 + 2 \eta\right)/3$ in
the same approximation, where the next-to-leading-order term is crucial to reflect
how small $\Delta m^2_{21}$ is as compared with $\Delta m^2_{31}$, because both
of them originate from proper breaking of the original exact neutrino mass degeneracy
in a flavor cube.
\end{itemize}
In short, a positive test of the correlation among $\varepsilon^{}_\xi$,
$\varepsilon^{}_\zeta$ and $\eta$ to a sufficiently good degree of accuracy in
the forthcoming JUNO, DUNE and Hyper-Kamiokande neutrino oscillation experiments
may serve as a smoking gun for the validity of our ansatz, provided the neutrino
mass spectrum turns out to be really normal and nearly degenerate.

For the sake of a numerical illustration, let us simply take
$\varepsilon^{}_\xi \simeq 1.8^\circ$ and $\eta \simeq 2.96 \times 10^{-2}$
to estimate the size of $\varepsilon^{}_\zeta$ with the help of Eq.~(\ref{9}).
We obtain $\varepsilon^{}_\zeta \simeq 2 \hspace{0.05cm} \varepsilon^{}_\xi
\simeq 3.6^\circ$, characterizing the deviation of $\theta^{}_{23}$ from
its $\mu$-$\tau$ symmetry limit (i.e., $\zeta^{}_* = 45^\circ$). Then
Eq.~(\ref{8}) leads us to the absolute neutrino mass scale
\begin{eqnarray}
\hspace{0.5cm}
m^{}_0 \simeq \sqrt{\frac{\Delta m^2_{21} + \Delta m^2_{31}}
{2\sqrt{2} \hspace{0.08cm} \varepsilon^{}_\xi}} \simeq 0.17 ~{\rm eV}
\; ,
\label{10}
\end{eqnarray}
after the central values of $\Delta m^2_{21}$ and $\Delta m^2_{31}$ listed
in Eq.~(\ref{4}) are input. This example
is hopefully helpful to illustrate the required precision levels of $\eta$,
$\varepsilon^{}_\xi$ and $\varepsilon^{}_\zeta$ that can be used to test
Eq.~(\ref{9}).

As a straightforward by-product of the above simple neutrino cuboid ansatz,
we proceed to give some brief discussions about its two non-oscillation
consequences as follows.
\begin{itemize}
\item     The sum of three neutrino masses, which can be well constrained
from a number of cosmological observations, is found to be
\begin{eqnarray}
\Sigma^{}_\nu \equiv m^{}_1 + m^{}_2 + m^{}_3 \simeq \sqrt{3} \hspace{0.05cm}
m^{}_0 \; ,
\label{11}
\end{eqnarray}
from Eq.~(\ref{7}), where the next-to-leading-order contributions have
all been cancelled out. A latest analysis based on the baseline $\Lambda$CDM
model has reported $\Sigma^{}_\nu < 64.2 ~{\rm meV}$ at the $95\%$ confidence
level~\cite{Jimenez:2026ycn}, which seems too stringent to support the normal and
nearly degenerate neutrino mass ordering. Note that current cosmological constraints
on $\Sigma^{}_\nu$ involve some undetermined uncertainties, and hence whether a
near neutrino mass degeneracy has been ruled out or not remains an open issue.
As pointed out in Ref.~\cite{Capozzi:2025wyn}, $\Sigma^{}_\nu < 0.2 ~{\rm eV}$
at the $2~\sigma$ level (within a factor of 3) is expected to be a reasonable
summary of those well-motivated bounds extracted from cosmology. In this case,
a normal and nearly degenerate neutrino mass spectrum is certainly allowed
\footnote{In fact, an approximate but instructive estimate tells us that
$\Sigma^{}_\nu \gtrsim 3\sqrt{\displaystyle\Delta m^2_{31}} \simeq 0.15 ~{\rm eV}$
is minimally required to assure the neutrino mass spectrum to be nearly degenerate.}.

\item     The effective electron neutrino mass term of the beta decays
can be expressed as
\begin{eqnarray}
\langle m \rangle^{}_\beta
\hspace{-0.2cm} & \equiv & \hspace{-0.2cm}
\sqrt{m^{2}_1 c^2_{12} c^2_{13} + m^{2}_2 s^2_{12} c^2_{13} + m^{2}_3 s^2_{13}}
\hspace{0.4cm}
\nonumber \\
\hspace{-0.2cm} & \simeq & \hspace{-0.2cm}
\frac{1}{\sqrt 3} \left(1 - \frac{1}{\sqrt 2} \hspace{0.08cm} \varepsilon^{}_\xi
- \frac{1}{3} \hspace{0.05cm} \varepsilon^{}_\zeta\right) m^{}_0 \; ,
\label{12}
\end{eqnarray}
implying $\Sigma^{}_\nu \simeq 3 \langle m\rangle^{}_\beta$ in the leading-order
approximation. An upper bound $\langle m\rangle^{}_\beta < 0.5 ~{\rm eV}$ at
the $2~\sigma$ level has been recommended in Ref.~\cite{Capozzi:2025wyn} after a
careful analysis of the KATRIN data~\cite{KATRIN:2024cdt}.
\end{itemize}
In comparison, the effective neutrino mass term
$\langle m\rangle^{}_{2\beta} \equiv \left|m^{}_1 c^2_{12} c^2_{13}
e^{{\rm i} 2\rho} + m^{}_2 s^2_{12} c^2_{13} e^{{\rm i} 2\sigma} +
m^{}_3 s^2_{13} e^{-{\rm i} 2\delta}\right|$ for the lepton-number-violating
neutrinoless double-beta decays is sensitive to the unknown CP-violating phases
$\rho$ and $\sigma$ and thus may involve some significant cancellation effects
for a normal neutrino mass spectrum no matter whether it is nearly degenerate
or not (see, e.g., Ref.~\cite{Cao:2019hli} for a detailed analysis).

In summary, we have introduced a neutrino cuboid to geometrically describe
the normal neutrino mass spectrum and shown that it can
accommodate both neutrino mass degeneracy and tribimaximal flavor
mixing in its cubic limit. This picture is motivated by the conjecture that
the observed large lepton flavor mixing phenomena may imply a peculiar neutrino
mass spectrum which is intrinsically different from the mass spectra of those
fundamental charged fermions. To pursue maximal experimental observability
and theoretical nontriviality, we have proposed a viable and testable
ansatz which predicts a normal but nearly degenerate neutrino
mass spectrum and a nearly tribimaximal neutrino mixing pattern. Some
phenomenological consequences of this ansatz has also been discussed.

\vspace{0.6cm}

The author was indebted to Ye-Ling Zhou for his warm hospitality at
HIAS in Hangzhou, where some aimless calculations were done and two
interesting preprints regarding the JUNO result were
seen~\cite{Ge:2026hsh,Chen:2026mvi}. He was also grateful to Di Zhang
for some helpful discussions during that period, although those
effects did not converge toward a nontrivial idea; and for his enlightening
comments on the present manuscript. This work was supported in part by the
National Natural Science Foundation of China under Grant No. 12535007 and
by the Scientific and Technological Innovation Program of the Institute of
High Energy Physics under Grant No. E55457U2.

\end{document}